\documentclass[aps,twocolumn,superscriptaddress,altaffilletter,lengthcheck,showpacs,showkeys]{revtex4}

\usepackage{amsmath,amssymb}      
\usepackage[dvips]{graphicx}
\usepackage[dvips]{graphicx}
\usepackage[latin1]{inputenc}
\usepackage[english]{babel}
\usepackage[dvipdf]{epsfig}
\usepackage{epsfig}
\newcommand{\be}{\begin{equation}}
\newcommand{\ee}{\end{equation}}
\newcommand{\bea}{\begin{eqnarray}}
\newcommand{\eea}{\end{eqnarray}}

\newcommand{\bphi}{\mbox{\boldmath $\phi$}}

\newcommand{\balpha}{\mbox{\boldmath $\alpha$}}

\newcommand{\bc}{\mathbf c}

\newcommand{\ben}{\begin{eqnarray}}
\newcommand{\een}{\end{eqnarray}}
\newcommand{\n}{\label}
\newcommand{\no}{\noindent}
\newcommand{\ga}{\gamma}

\usepackage{color}

\begin{document}

\title{Internal space structure  generalization of the quintom cosmological scenario}


\author{Luis P. Chimento}
\email{chimento@df.uba.ar}
\affiliation{Departamento de F\'isica,
Universidad de Buenos Aires, 1428 Buenos Aires, Argentina}
\author{M\'onica Forte}
\email{monicaforte@fibertel.com.ar}
\affiliation{Departamento de F\'isica,
Universidad de Buenos Aires, 1428 Buenos Aires, Argentina}
\author{Ruth Lazkoz}\email{ruth.lazkoz@ehu.es}
\affiliation{Fisika Teorikoa, Zientzia eta Teknologia Fakultatea, 
Euskal Herriko Unibertsitatea, 644 Posta Kutxatila, 48080 Bilbao, Spain}
\author{Mart\'{i}n G. Richarte}\email{martin@df.uba.ar}
\affiliation{Departamento de F\'isica,
Universidad de Buenos Aires, 1428 Buenos Aires, Argentina}

\date{\today}

\begin{abstract}
We introduce the Lagrangian for a multi-scalar field configuration in a $N$-dimensional internal space endowed with a constant metric $Q_{ik}$ and generalize the quintom cosmological scenario. We find the energy momentum tensor of the model and show that the set of dual transformations, that preserve the form of the Einstein equations in the Friedmann-Robertson-Walker (FRW) cosmology, is enlarged. We show that the stability of the power law solutions leads to an exponential potential which is invariant under linear transformations in the internal space. Moreover, we obtain the general exact solution of the Einstein-Klein-Gordon equations for that potential. There exist solutions that cross the phantom divide and solutions that blow up at a finite time, exhibiting a superaccelerated behavior and ending in a big rip. 
We show that the quintom model with a separable potential can be identified  with a mixture of several fluids. This framework  includes the $\Lambda$CDM model, a bouncing model, and a setting sourced by a cosmic string network plus a cosmological constant. The we concentrate on the case where the dimension of the internal quintessence sector $N_{q}$ exceeds the dimension of the internal phantom sector $N_{ph}$. For $(N_q,N_{ph})=(2,1)$ the dark energy density derived from the 3-scalar field crosses the phantom divide and its negative component plays the role of the negative part of a classical Dirac Field.

\end{abstract}

\maketitle


\section{Introduction}


A large number of recent cosmological data, including  Type Ia supernovae
\cite{Ries:2004}, Large Scale Structure \cite{Tegmark:2004,morelss} and Cosmic
Microwave Background anisotropies \cite{wmap}
have provided  strong evidences for a spatially flat and accelerated
expanding universe at the present time. For general relativistic
FRW cosmologies, such an accelerated expansion implies the existence of a mysterious exotic matter with large enough negative pressure, the dark energy, whose energy density has been dominating the recent stage of the universe.
The astrophysical feature of dark energy is that it remains unclustered at
all scales where gravitational clustering of baryons and
nonbaryonic cold dark matter can be seen. The combined analysis of the different
cosmological observations suggests that the universe consists of about 70\%
dark energy, 30\% dust matter (composed by cold dark matter and baryons),
and negligible radiation. Although the nature and the origin of dark energy are unknown,
several candidates have been proposed to describe it. The  simplest theoretical candidate
is the cosmological constant $\Lambda$ \cite{Lambda1}, \cite{Lambda2} whose equation of state is $p=-\rho$.
However, this proposal suffers from the well-known cosmological problem according to which
there is a unexplained extraordinary discrepancy (of about 120 orders of magnitude) between
the observed value of $\Lambda$ and the value predicted by quantum field theory \cite{Steinhardt:1997}, \cite{Copeland:2006}.
For this reason, alternatives routes have been suggested to study the dark energy and its dynamical
evolution \cite{Copeland:2006}. So far a wide variety of scalar-field dark energy models have been studied,
including quintessence \cite{Quintaesencia}, $k$-essence \cite{Kesencia}, tachyon \cite{Taquion}, phantom \cite{Fantasma}, 
ghost condensate \cite{FantaCondensados} and more recently, quintom dark energy \cite{quintom, QuintomRuth, StringQuintom}.
In addition, other possibilities include interacting dark energy \cite{InteracDE}, braneworld models \cite{Branas},
Chaplygin gas models \cite{Chaplygin}, holographic dark energy \cite{HoloDE} and many others.

The quintom model is a hybrid with a quintessence component, usually constructed by a scalar field
and a phantom scalar field minimally coupled to gravity. This model has aroused interest due to a debate about the possibility that
the dark energy equation of state crossed the phantom divide in the near past.
Notice that in the quintessence models the kinetic energy is always positive definite for a positive definite potential, whereas
one has exactly the opposite in phantom models. So, neither the quintessence nor the phantom alone can fulfill the transition a non-phantom to phantom stage. 

Many authors have studied the quintom paradigm with several potentials from the dynamical autonomous system perspective \cite{CaiWei:2005DS},\cite{QuintomRuth}. In this context, some of them have performed a general analysis by including  the contribution of both a quintom component (dark energy) and a matter component into the Friedmann equation. For example, in  Ref.\cite{QuintomRuth} it was found that the phantom attractor is not generic, which means that in a quintom model there may exist either de Sitter attractors associated with the saddle points of the potential, or tracking attractors in the asymptotic regime where the scalar fields diverge.

Finally, we want to stress that motivated by string theory a quintom scenario has been found by taking into account the non-perturbative effects of a generalized Dirac-Born-Infeld action \cite{StringQuintom}. Moreover, a ``No-Go Theorem" has been proven to constrain the building of quintom scenario \cite{TeoremaNoGo}. According to this theorem in order to achieve the crossing of the phantom divide it is necessary to have more than one degree of freedom. Clearly, this theorem offers a concrete theoretical justification for the quintom paradigm. 

In this paper we present a multi-scalar field model embedded in a $N$-dimensional 
internal space structure endowed with a constant internal metric $Q_{ik}$.
In Sect. II, we introduce the Lagrangian for the multi-scalar field model, calculate the energy momentum tensor and obtain the energy density and  pressure of the multi-scalar field. We also study how the dual transformation, that links contracting and expanding cosmologies is realized when the 
internal structure is taken into account, and we then perform the stability analysis for power law solutions. In Sect. III , we report the general exact solution of the Einstein-Klein-Gordon equations for the exponential potential and illustrate the existence of  the crossing of the phantom divide, showing that the model has a smooth transition from the quintessence to the phantom era. 
In Sect. IV we study quintom and a 3-dimensional scalar field models driven by  separable potentials. Finally, in Sec. V the conclusions are stated. Throughout this paper we shall use natural units ($G=c=1$). 

\section{Multi-scalar field cosmologies}

We wish to construct a general class of dark energy models with  
late-time phantom behavior motivated by the fact that this peculiar 
sort of expansion fits confortably within current observational results. 
Accelerated expansion cannot extend indefinitely back in time, because 
it would spoil the formation of structures, 
and well founded theoretical arguments support  the view that our 
universe was dark matter dominated for a long time span before it entered 
the current dark energy dominated accelerated phase. In addition, 
results point out in the direction that dark energy at present does not 
just produce acceleration but what is rather called super-acceleration, 
thus indicating that dark energy is phantom-like, and it is quite likely that
it was it actually became phantom-like just recently,
i.e. dark energy may have crossed the so-called phantom divide.

The mainstream approach (for simplicity and because there is no 
evidence in favor of the contrary)  is to consider that dark energy and dark 
matter interact only gravitationally, so they ``see'' each other 
by their effects in the expansion. Intuition suggests, then, that  the 
ability for a dark energy fluid to cross the divide is not ruined by the 
presence of dark matter. For that reason, even if one's goal is to try  
and construct cosmological scenarios containing dark matter and dark 
energy featuring a crossing of the phantom divide, it makes sense to start 
off by considering the dark energy fluid alone. Put another way, it is 
very likely that the construction of a dark energy source with the 
desired property leads when combined with dark matter to a cosmological 
scenario in which the dark energy bit retains the same feature.

Quite a few scalar field model configurations with the ability of 
crossing the phantom divide can be found in the literature, and perhaps the 
most populated class is that of  the generically dubbed quintom models \cite{quintom, QuintomRuth, StringQuintom}. 
Their main feature is that the kinetic energy is not a definite 
positive quantity. Quintom scenarios have attracted considerable attention 
because the mentioned property of the kinetic energy allows giving a 
unified treatment of phantom and conventional scalar fields. 

Taking quintom models as an inspiration we assume a $N$-dimensional 
internal space structure endowed with a constant metric $Q_{ik}$ and propose a 
new theoretical setup with multiple fields $\phi_{i}(x)$ where $i= 1...N$. 
This composite field can be thought of as a vector within the internal space.
For any two vectors $u_i$ and $w_k$ belonging to this space their scalar product is 
$\mathbf u\cdot \mathbf w\equiv Q_{ik}u_{i}w_{k}$
and we denote the norm as $\mathbf w\cdot \mathbf w\equiv w^2=Q_{ik}w_{i}w_{k}$.  
In what follows we use latin indexes for the internal space and 
summation over repeated indexes will be assumed. 
 
The Lagrangian for this model is given by 
 \begin{equation}
{\cal{L}}=
\frac{1}{2}Q_{ik}\partial_{\mu}\phi_{i}(x)\partial^{\mu}\phi_{k}(x) - V(\phi_1,\phi_2,...,\phi_N),
\label{nquinT}
\end{equation}
where $V(\phi_1,\phi_2,...,\phi_N)$ is a selfinteracting scalar potential depending on the fields $\phi_1,\phi_2,...,\phi_N$ only. The energy momentum $T_{\mu\nu}$ associated to  the multi-scalar field Lagrangian (\ref{nquinT}) is obtained through the standard equation $T_{\mu\nu}=2\delta {\cal{L}}/{\delta g^{\mu\nu}} - g_{\mu\nu}{\cal{L}}$, so we get 
\begin{equation}
T_{\mu\nu}=Q_{ik}\partial_{\mu}\phi_{i}\partial^{\mu}\phi_{k} -
g_{\mu\nu}\left[\frac{1}{2}Q_{ik}\partial_{\xi}\phi_{i}\partial^{\xi}\phi_{k}-V\right].
\end{equation}

Our framework is that of spatially flat FRW 
cosmologies and the energy density and pressure of the dark energy fluid are
\be
\rho=\frac{1}{2}Q_{ik}\dot{\phi}_{i}\dot{\phi}_{k}+V, \qquad 
p=\frac{1}{2}Q_{ik}\dot{\phi}_{i}\dot{\phi}_{k}-V,
\ee
where we have assumed homogeneous scalar fields $\phi_i=\phi_i(t)$. The evolution of a universe sourced by that configuration of multiple interacting fields in the fashion above is governed by the equations
\begin{eqnarray}
\n{00}
&&3H^2=\frac{1}{2}Q_{ik}\dot{\phi}_{i}\dot{\phi}_{k}+V \label{ein00},\\
\n{11}
&&\ddot{\phi}_{i} + 3H \dot{\phi}_{i}+Q^{-1}_{ik}\frac{\partial V}{\partial\phi_{k}}=0, \label{KGgral}
\end{eqnarray}
and they get combined to give 
\begin{equation}
\dot H=-\frac{1}{2}Q_{ik}\dot{\phi}_{i}\dot{\phi}_{k}\label{ein11}.
\end{equation}
Our proposal is to use this extension which, due to the presence of an 
internal space, allows to consider a richer structure than in 
the precursor quintom cosmologies. The usual quintom model fits 
in our description for the choice $Q_{ik}={\rm diag}~(1,-1)$. 
In particular the internal metrics $Q_{ik}={\rm diag}~(1,0)$ and $Q_{ik}={\rm diag}~(0,-1)$ allow to reproduce to quintessence and phantom cosmologies respectively. 

\subsection{Duality}

Interestingly, the internal space structure also allows to interpret 
duality transformations linking contracting and expanding conventional 
cosmologies in a very intuitive and elegant fashion. In fact, the $a\to 
1/a\,$ duality transformation implies $H\to-H$, and from the Einstein
equations (\ref{ein00}) and (\ref{ein11}) it follows that there are two 
alternative ways to achieve the transformation that leaves  those 
equations unaltered: 
\begin{itemize}
\item Case 1
\begin{eqnarray}
&&\dot\phi_k\to i\dot\phi_k, \quad V\to 
Q_{ik}\dot\phi_i\dot\phi_k+V(\phi_i)
 \end{eqnarray}
\item Case 2
\begin{eqnarray}
&&Q_{ik}\to -Q_{ik}, \quad V\to Q_{ik}\dot\phi_i\dot\phi_k+V(\phi_i)
 \end{eqnarray}
\end{itemize}
A realization of the duality as given by case 1) requires  potentials 
which remain real under the transformation of the fields $\dot\phi_k\to 
i\dot\phi_k$. However, the realization induced by the internal 
structure and given by case 2) is of wider applicability as the former 
requirement on the potential is not necessary. In addition, both the 
transformed and non-transformed fields are real.
For instance, phantom cosmologies characterized by a future singularity 
occurring in finite time can be obtained by performing the $a\to 1/a$ 
duality transformation onto cosmologies with a final big crunch. For the 
case 1), one of them (phantom or big crunch cosmology) will be 
associated with imaginary fields. However, for the case 2) both cosmologies are 
described by real fields. 

\subsection{Stability of power law solutions}

We are going to investigate now the existence of 
asymptotic power law solutions by means of a structural stability analysis, in 
other words, we are going to unveil the asymptotic nature of potentials 
allowing such stable solutions.  To this end we introduce the 
barotropic index $\gamma=(\rho+p)/\rho$, and  from the definitions above and the 
Einstein equations (\ref{ein00}) and (\ref{ein11}) one gets the dynamical 
equations for $\gamma=-2 \dot H /3 H^2$, namely
\begin{equation}
\dot \gamma=(\gamma-2)\left(3H\gamma+\displaystyle\frac{\dot 
V}{V}\right) \label{gammadot}.
\end{equation}
Clearly, the solution $\gamma=2$ represents an equilibrium point, but 
it corresponds to non-accelerated expansion, specifically, $a\propto 
t^{1/3}$. Given that practically any other interesting cosmological model one can 
think of will have $\gamma<2$, let us demand then that Eq. (\ref{gammadot}) 
admits another equilibrium point with $\gamma\equiv\gamma_0$  representing 
in consequence a solution with $a\propto t^{2/3\gamma_0}$. In 
particular, a solution representing a phantom late-time attractor would be 
characterized by a constant and negative value of $\gamma$. Below, we will show 
that the requirement of the existence of such an equilibrium point 
restricts the functional form of the $V(\phi_i)$ potential. Then we 
formulate a structural stability analysis 
of  the equation governing the evolution of the parameter $\gamma$. 
This is  equivalent to imposing the asymptotic condition on the 
potential $\dot V+3\gamma_0HV=0$, which can be integrated to give
$V=V_0a^{-3\gamma_0}$ with $V_0$ a positive integration constant.  
By choosing this potential the asymptotic regime of $\gamma$ is governed by the equation
\begin{equation}
\dot \gamma=3H(\gamma-2)(\gamma-\gamma_0)\label{newgammadot}.
 \end{equation}
Let us now  define $\gamma=2-\epsilon$ and recast Eq. 
(\ref{newgammadot})  in terms of this new definition. By expanding the rhs of this 
equation up to order $\epsilon$, for expanding universes ($H>0$), the 
solution $\gamma=2$ is a repeller. A similar analysis can be done 
for the case $\gamma=\gamma_0\ne 0$ (the particular case $\gamma_0=0$ 
will investigated at the end of this section). If one defines
  $\gamma=\gamma_0-\epsilon$ it turns out that the $\gamma=\gamma_0$ 
solution is  an attractor provided $\gamma_0<2$, which is on the other 
hand our working hypothesis. If one then calculates $V$ and $H$ as 
functions of $t$ for   $a= t^{2/3\gamma_0}$ it can be inferred from Eqs. 
(\ref{ein00}) and (\ref{ein11}) that  
\begin{equation}
 V=\frac{V_0}{t^2},  \qquad V_0=\frac{2(2-\gamma_0)}{3 \gamma_0^2}.
\end{equation}

As it has been shown, in spite of having started from a general 
formulation, we have been able to calculate the value of $V_0$ and the 
asymptotic dependence of the potential on cosmic time. However, in what the 
reconstruction of the potential is concerned we need to make contact with  standard 
formulations in physical cosmology and to specify the kinetic energy as 
a function of the fields and their derivatives.

For the attractor solution (which satisfies a power law) it can be seen 
that a satisfactory choice for the asymptotic behavior of the scalar 
fields is $\bphi=\bphi_{0}\log t$
with $\bphi_{0}$ a constant vector in the internal space. Besides, we 
are considering that physical quantities such as the potential energy 
depend only on scalar quantities in this internal space, so for 
simplicity we will consider the potential depends on the scalar fields through a 
linear form of the fields. To that end we introduce a constant vector 
$\balpha$ in the internal space. Under these assumptions we build the scalar quantity $\bphi=\balpha\cdot\bphi_0\log t$,
which leads to 
\begin{equation}
V=V_0e^{-\balpha\cdot \bphi}\label{exppot},
\end{equation}
where we have fixed  $\balpha\cdot\bphi_0=2$.  One can also find the 
following relations 
\begin{eqnarray}
 \bphi_0=\frac{{2}\balpha}{\alpha^2},\qquad\alpha^2=3\gamma_0.
\end{eqnarray}
Finally,  in terms of the parameters of our framework the attractor 
solution gets recast as
\begin{equation}
a=t^{2/\alpha^2}.
\end{equation}
Due to the internal space structure of our model the exponent of the 
latter can take either sign, so both phantom and non-phantom power law 
solutions are available.

Finally, we conclude this section with the investigation of  the 
stability of the solution $\gamma_0=0$. If  we impose the fulfillment of  Eq. 
(\ref{gammadot}) asymptotically one obtains  $\dot V/V=0$ with $V=V_0$ a 
constant. Now defining $\gamma=-\epsilon$ and using again 
(\ref{newgammadot}) it follows that de Sitter solution ($H=H_0$) is asymptotically 
stable.

\section{The exponential potential}

We have investigated the stability of the power law solution
with a exponential potential and the possibility of phantom and no phantom behavior for the scale factor.
In this section we shall find the general exact solution for our $N$-scalar field model driven by the exponential potential (\ref{exppot}) in the spatially flat FRW spacetime. In this case the Eqs. (\ref{ein00})-(\ref{KGgral})  
represent a non-linear coupled system of $(N+1)$ differential equations for the scale factor $a(t)$ 
and the $N$ scalar fields $\phi_{i}(t)$. For a non-phantom scalar field driven by a exponential potential and a free scalar field, the general solution of the system of equations (\ref{ein00})-(\ref{KGgral}) was found in  Ref. \cite{Soluc.Exacta}. Using the Einstein equation
(\ref{ein11}) it is easy to prove that the geometrical object
\begin{equation}
\dot{\phi_{k}}= \alpha_{k}H + \frac{c_{k}}{a^{3}}\label{integralp},
\end{equation}
with $c_{k}$ a constant vector in the internal space, is a first integral of Eq. (\ref{KGgral}). Inserting the latter into the Eq. (\ref{ein11}) and expanding, we obtain
\begin{equation}
-2\dot{H}=\alpha^2H^{2}+2\balpha\cdot\bc\,\frac{H}{a^{3}} +\frac{c^2}{a^{6}}.\label{dobleprod}
\end{equation}
Making the change of variables 
\begin{equation}
s=a^{-3/n}, \qquad \tau=\balpha\cdot\bc\,\,t, \qquad n=-\frac{6}{\alpha^2},\label{cambioent}
\end{equation}
in Eq. (\ref{dobleprod}), it becomes a nonlinear second order differential equation for $s(\tau)$:
\begin{equation}
\ddot s + s^{n}\dot s+\frac{1}{4\cos^2{\theta}}\,s^{2n + 1}=0,\label{NL}
\end{equation}
where the dot stands for differentiation with respect to the argument of the function and
\begin{equation}
\cos{\theta}=\frac{\balpha\cdot \bc}{\alpha c}\label{def.b}.
\end{equation}
The sign of the parameter $n$ depends exclusively of the internal metric $Q_{ik}$. In particular, for an Euclidean metric this parameter becomes negative. 
After the new variable $s$ is calculated,
we obtain the scale factor $a$ in terms of the cosmic time $t$ by using Eqs. (\ref{cambioent}).

Inserting the first integral (\ref{integralp}) into the Friedmann equation (\ref{ein00}), we 
get a quadratic equation in the expansion rate $H$ which has real solutions only when its discriminant
is positive definite, this condition can be recast as $
(\balpha\cdot \mathbf c)^{2}+(6-\alpha^2)(c^2 + 2a^{6}V)>0$.
It gives a relation among the integration constants and the exponent of the exponential potential.

Eq. (\ref{NL}) can be linearized and solved using the  form invariance group of
this nonlinear differential equation under a non local change of variable. This technique relates a class of nonlinear differential 
equations with the damped harmonic oscillator one. In fact, Eq. (\ref{NL}) belongs to the following  general class of second order nonlinear ordinary 
differential equations:
\begin{equation}
\ddot{s} + \mu~F(s)\dot{s}+ \nu~F(s)\int{F(s)ds}+\delta~F(s)=0\label{IF1}, 
\end{equation}
where $s=s(t)$, $F(s)$ is a real arbitrary function, 
and $\mu$, $\nu$, $\delta$ are constant parameters.

By using the new pair of variables $(h,\eta)$, defined by the non-local transformation 
\begin{eqnarray}
F(s)ds=G(h)dh,\n{n1}
\qquad 
F(s)dt=G(h)d\eta, 
\end{eqnarray}
Eq.(\ref{IF1}) becomes the nonlinear ordinary differential equation
\begin{equation}
h''+\mu G(h)h' + \nu G(h)\int{G(h)dh}+\delta~G(h)=0\label{IF2}, 
\end{equation}
where $'$ denotes differentiation with respect to $\eta$. Interestingly, Eqs. (\ref{IF1}) and (\ref{IF2}) are related between them by means of the formal 
change $F\leftrightarrow G$ and $s\leftrightarrow h$. Then, the non-local transformation (\ref{n1}) preserves the form of these equations mapping solutions of  Eq. (\ref{IF1}) into solutions of  Eq. (\ref{IF2}) for any function $F(s)$ and $G(\eta)$, thus linking solutions of 
two different physical configurations. Taking into 
account the nature of this non-local change of variables it is not always possible
to find explicit solutions.
However, we can use the form invariance to relate solutions of a nonlinear equation (\ref{IF1}) 
with a linear one. In our case, choosing the specific parametrization $F(s)=s^{n}$ and $G(h)=1$ in the 
change of variables (\ref{n1})
\begin{eqnarray}
h=\int{s^{n}ds},\n{c1}
\qquad
\eta=\int{s^{n}dt}.
\end{eqnarray}
and introducing these new variables in Eqs. (\ref{IF1}) and (\ref{IF2}), they become
\ben
\ddot{s}+ \mu~s^{n}\dot{s}+ \frac{\nu}{n+1}\,s^{2n+1}+\delta s^{n}=0,\n{s}\\
h''+ \mu h' + \nu h+\delta=0. \n{h}
\een
It shows that the non-local change of variable (\ref{c1}) transforms the non-linear differential equation (\ref{s}) into the linear damped  harmonic oscillator equation (\ref{h}). 

Comparing our equation (\ref{NL}) with Eqs. (\ref{s}) and (\ref{h}), for $\mu=1$, we have two different cases to analyze according to the $n$ value,

\begin{itemize}
	\item $n\neq -1$ and $\delta=0$ ~$\Rightarrow~F(s)=s^{n},~~~~~\nu=\frac{(n+1)}{4\cos^2\theta}$
	\item $n~= -1$ and  $\nu=0$ $\Rightarrow~ F(s)=\frac{1}{s},~~~~~~\delta=\frac{1}{4\cos^2\theta}$
\end{itemize}
Below we find the general solution of the Einstein-Klein-Gordon equations for the exponential potential.

\subsection{Explicit solution for $n\neq -1, -2$ and $\nu=\frac{(n+1)}{(n+2)^{2}}$}


In this case explicit $\tau$-dependent solutions can be found introducing the function $v(\tau)$ (for more detail see Ref. \cite{Soluc.Exacta}) and making the  substitution 
\begin{equation}
s^{n}=\left(\frac{n+2}{n}\right)\frac{v^{n}}{k_{1} + \int{v^{n}d\tau}},\label{sustitu}
\end{equation}
into the equation (\ref{NL}). This  reduces it to $\ddot{v}=0$, whose solution is $v(\tau)=k_{2} + k_{3}\tau$. After inserting the solution $v$ in  Eq.(\ref{sustitu}), integrating and using $s(a)$ as given by Eq.(\ref{cambioent}), we  obtain the scale factor 
\begin{equation}
a(v)=\left[{\frac{n}{(n+1)(n+2)}}\right]^{1/3}\left[ v + k|v|^{-n}\right]^{1/3},\label{a(v)}
\end{equation}
where the constant $k$ has been expressed in terms of the old one\footnote[2]{The constant $k_{3}$ can be set equal to the unity scaling the time variable. Also, we choose 
$k_{2}=-\tau_{0}$, then $v=\Delta\tau=\tau -\tau_{0}$.}.
Finally, we can get the components $\phi^{i}$ of the multi-scalar field $\bphi$ integrating  the Eq. (\ref{integralp})  
\begin{equation}
\frac{d\phi_{k}}{dv}=\alpha_{k}\frac{d\ln{a}}{dv} + \frac{c_{k}a^{-3}}{\balpha\cdot \bc}.\label{integralp2}
\end{equation}
To this end, we rewrite the second term of the Eq. (\ref{integralp2}) using the definition given in  Eq. (\ref{cambioent}) as
\begin{equation}
a^{-3}=\frac{n+2}{n}\frac{d}{dv}\left[\ln{\left(k_{1} + \int{v^{n}d\tau}\right)}\right].\label{sgtermino}
\end{equation}
Inserting the latter into the Eq. (\ref{integralp2}) we obtain the $\phi^{i}(v)$
\begin{equation} 
\n{fi}
\Delta\phi_{i}=\alpha_{i}\ln{|a|}
+\frac{(n+2)c_{i}}{n(\balpha\cdot\bc)}\ln{\left|\frac{n+2}{n}\,v^{n}a^{3}\right|},
\end{equation}
where $\Delta\phi_{i}(v)=\phi_{i}(v)-\phi_{0i}$ and $\phi_{0i}$ are $N$ integration constants \footnote[3]{Without loss of
generality we can choose $\phi_{0i}\rightarrow \frac{\alpha_{i}}{\alpha ^2}log|\phi_{0}|$ where $\phi_{0}$ is an arbitrary constant.}. For positive $n$ , when $v\rightarrow 0$, the leading terms in the scale factor is given by
\begin{equation}
a\propto k |v|^{-\frac{n}{3}} ,\label{alterm} 
\end{equation}
So, the scale factor blows up near the origin $v=0$ and the components of the  multi-scalar field have a logarithmic divergence $\Delta\phi_{i} \propto \ln|v|$ for all $n$. We find a resemblance  
with the big rip behavior in the phantom universe reported recently by 
Wei \cite{Weibigrip:2005} with an inverse power potential $V=V_{0}\phi^{-1}$.

As we have integrated the second order differential for the scale factor (\ref{ein11}), instead of the original Friedmann equation (\ref{ein00}), we have to constrain the integration constants. By using Eqs.(\ref{ein00}) and (\ref{ein11}), we obtain $3H^{2}+\dot{H}=V$, which can be rewritten in terms of the variable $u= a^{3}$ as
\begin{equation}
\frac{d^{2}u}{dv^{2}}=\frac{3uV}{(\balpha\cdot\bc)^{2}}.\label{sumaFriedHP2}
\end{equation}
After replacing the solutions $u(v)$ and $\phi(v)$ in Eq.(\ref{sumaFriedHP2}) the dependence in the variable
$v$ is missed and we obtain the following relation between the integration constants
\begin{equation}
k=-\frac{V_{0}}{2c^2|\phi_{0}|\cos^{2}{\theta}}~sign\left(\frac{n+2}{n}\right).\label{consistencia}
\end{equation}
Hence, the scale factor (\ref{a(v)}) and the multi-scalar field (\ref{fi}) whose integration constants verify the relation (\ref{consistencia}) are exact solutions of the Einstein-Klein-Gordon equations (\ref{ein00})-(\ref{KGgral}).

\subsection{Explicit solution for $n= -1$ and $\nu=0$}

In the $n=-1$ case, following the same steps we get 
\begin{equation}
a=\left[|v||b_{1}+ \ln{|v|}|\right]^{1/3},\label{anm1}
\end{equation}
\begin{equation}
\Delta\phi_{i}=\frac{\alpha_{i}}{3}\ln{|v||b_{1}+\ln|v||}-\frac{c_{i}}{\balpha\cdot\bc}\ln{|b_{1}+ \ln|v||}\label{camponm1},
\end{equation}
where $b_1$ is an integration constant. Inserting Eqs. (\ref{anm1})-(\ref{camponm1}) into the Eq. (\ref{sumaFriedHP2}) 
we find the relation $V_{0}=2|\phi_{0}|c^2\cos^{2}{\theta}$, where $\alpha^2=6$ and the constant $\phi_{0}$ has been renamed properly. 

\subsection{Implicit solution}
The general implicit solution of Eq. (\ref{NL}) can be found 
for arbitrary choices of the parameters by
solving Eq. (\ref{h}) and using Eqs. (\ref{cambioent}), (\ref{c1}). In this case we have two different set of solutions according to the value of $n$. For $n\neq-1$, $\delta=0$ and  $\mu=1$ the scale factor is given by
\begin{equation}
a(\eta)=\left[ \sqrt{(n+1)}\left(b_{1}\exp{\lambda_{-}\eta}+ b_{2}\exp{\lambda_{+}\eta}\right)\right]^{-n/3(n+1)},\label{aparamet}
\end{equation}
where $\lambda_{\pm}$ are the roots of the characteristic polynomial of the Eq. (\ref{h}), $\lambda_{\pm}=(-1 \pm \sqrt{1-4\nu})/2$. Also, from Eq. (\ref{integralp}) we calculate the multi-scalar field 
\begin{equation}
\n{pi}
\phi_{i}(\eta)=\phi_{i0}+\alpha_{i}\ln a(\eta) + \frac{c_{i}}{\balpha\cdot\bc}\eta,
\end{equation}
where $\phi_{i0}$ are integration constants. Now, inserting Eqs. (\ref{aparamet}) and 
(\ref{pi}) into the Friedmann equation (\ref{sumaFriedHP2}) the integration constants are constrained to 
\begin{equation}
b_{1}b_{2}=\frac{V_{0}}{2(1-4\nu)c^2|\phi_{0}|\cos^{2}{\theta}}.\label{consistenciaparam1}
\end{equation}

Finally, in the $n= -1$, $\nu=0$ and $\delta=(4 \cos^{2}{\theta})^{-1}$ case we have
\begin{eqnarray}
&&a(\eta)=\exp{\left[\frac{1}{3}\left(b_{2}-\delta\eta+b_{1}e^{-\eta}\right)\right],\label{aparamet2}}
\\
&&\phi_{i}(\eta)=\phi_{i0}+\alpha_{i}\ln a(\eta) + \frac{c_{i}}{\balpha\cdot\bc}\eta, \label{phiparame2}
\\
&&b_{1}=\frac{V_{0}}{2|\phi_{0}|c^2\cos^{2}{\theta}} \label{consparame2}.
\end{eqnarray}
For $\theta=0$ we reobtain the standard formula
for two scalar fields driven by an exponential potential Ref. \cite{Soluc.Exacta}.

\section{$2,3$-dimensional internal space}

Below we investigate several interesting examples with different internal metrics.

\subsection{Big Rip solution}

Let us explore a new class of quintom model associated with the internal metric 
\begin{equation} 
Q_{ik}=
\begin{pmatrix}
0&1\\
1&0\
\end{pmatrix}.
\end{equation}
As the sign of the kinetic term is non definite, a new feature arises, the possibility of having a smooth transition between quintessence and phantom  scenarios (see Fig.1). 
\begin{figure}[!ht]
\centering
\includegraphics[height=6cm, width=8cm]{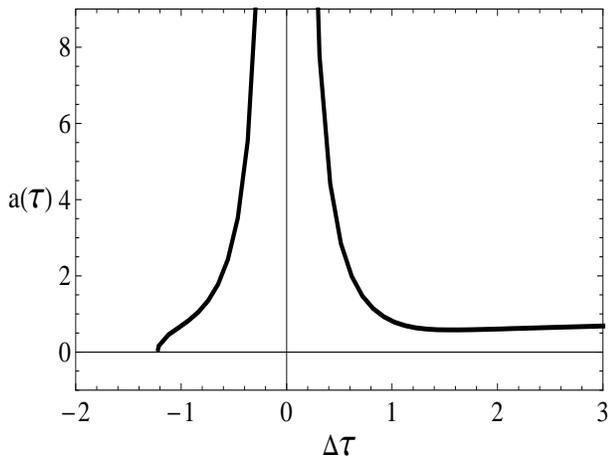}
\caption{The two branches of the scale factor for $\balpha^2=-1$.}
\end{figure}
For $\balpha^2<0$ the scale factor has two branches. In the branch $\Delta\tau<0$ the universe expands and has a finite time span, it begins to evolve from an initial singularity, has a transient phase to a superacelerated expanding scenario ending in a big rip. Initially the universe is dominated  by an quintessence 
scenario with kinetic energy $T(\phi_{1},\phi_{2})>0$ and barotropic index $\gamma=(p+\rho)/\rho>0$, after that, the universe evolves to a dominated phantom dark energy stage with
$T(\phi_{1},\phi_{2})<0$ and $\gamma<0$ until the big rip occurs when $\Delta\tau\rightarrow 0^{-}$ and  $a\rightarrow \infty ^{+}$. The point where the kinetic term changes its sign corresponds to the inflexion point of the scale factor (see Fig. 1).
The branch $\Delta\tau>0$  represents a bouncing universe with a non-vanishing absolute minimum.

\subsection{Quintom}

We consider the standard quintom model with internal metric $Q_{ik}=\rm {diag}(1,-1)$ and assume that the scalar fields $\phi_1$ and $\phi_2$ are coupled through the potential $V(\phi_{1},\phi_{2})$. So, the Friedmann equation reads
\begin{eqnarray} 
3H^{2}=\frac{1}{2}({\dot{\phi}}^2_{1}-{\dot{\phi}}^2_{2}) + V(\phi_{1},\phi_{2}),\label{fried2}
\end{eqnarray}
where $\phi_{1}$ represents the quintessence and $\phi_{2}$ plays the role of a phantom field. The system is closed with the Klein-Gordon equations, one for each field.
\begin{eqnarray}
\ddot{\phi}_{1} + 3H\dot{\phi_{1}} + \partial_{\phi_{1}}V(\phi_{1},\phi_{2})=0, \label{ec.kgreal}
\\
\ddot{\phi}_{2} + 3H\dot{\phi_{2}} - \partial_{\phi_{2}}V(\phi_{1},\phi_{2})=0. \label{ec.kgfantasma}
\end{eqnarray}
We propose a potential which can be written as follows,
\begin{equation}
V(\phi_{1},\phi_{2})=V_{0} + \frac{\sigma}{2}{\dot{\phi}}^2_{1}+ \frac{\beta}{2}{\dot{\phi}}^2_{2},\label{pot.gnral}
\end{equation}
where $\sigma$ and $\beta$ are constant and $V_{0}>0$. It can mimics a simple model containing a mixture of dark matter and dark energy where the former is the clustering component and the latter give rise to the current acceleration. With the help of the Eq. (\ref{pot.gnral}) we find that 
\be
\label{grad.1}
\partial_{\phi_{1}}V=\sigma\ddot{\phi}_{1}, \qquad 
\partial_{\phi_{2}}V=\beta\ddot{\phi}_{2},
\ee
where we have assumed separability of the potential $V(\phi_{1},\phi_{2})=V(\phi_{1})+V(\phi_{2})$. Thus, Eqs. (\ref{ec.kgreal})-(\ref{ec.kgfantasma}) are easily integrated 
\be
\label{aphipto.1}
\dot{\phi}_{1}=\dot{\phi}_{01}~a^{-3/(1+\sigma)}, \qquad 
\dot{\phi}_{2}=\dot{\phi}_{02}~a^{-3/(1-\beta)},
\ee
whit $\dot{\phi}_{01}$ and $\dot{\phi}_{02}$ integration constants.
Combining Eqs.  (\ref{fried2}), (\ref{pot.gnral}), (\ref{aphipto.1}), the Friedmann equation becomes
\begin{equation}
3H^{2}=V_{0}+\frac{1}{2}\frac{(\sigma+1){\dot{\phi}}^2_{01}}{a^{6/(1+\sigma)}}+
\frac{1}{2}\frac{(\beta-1){\dot{\phi}}^2_{02}}{a^{6/(1-\beta)}}. \label{Hdea}
\end{equation}
On the other hand, the effective barotropic index $\gamma=(\rho + p)/\rho$ of this mixture 
\be
\n{gm}
\ga=\frac{\dot{\phi}_{01}^2~a^{-6/(1+\sigma)}-\dot{\phi}_{02}^2~a^{-6/(1-\beta)}}{\rho},
\ee
shows the occurrence of a smooth transition from the no phantom regime to the phantom one. 
This transition is achieved  when $\ga(a_{ph})=0$, that is at $a_{ph}= (\dot\phi_{02}/\dot\phi_{01})^{(1-\beta)(1+\sigma)/3(\sigma+\beta)}$.


In the case $\sigma=1$ and $\beta=3$, we associate the second term of the Friedmann equation (\ref{Hdea}) with  an attractive component $\rho_m=\dot{\phi}^2_{01}/a^{3}$, which includes both baryonic and nonbaryonic matter, having the equation of state $p_{m}\approx 0$, while the remaining two terms constitute the dark energy component $\rho_{de}=V_{0}+\dot{\phi}^2_{02}a^{3}$. The Eq. (\ref{Hdea}) with $\dot{{\phi}}^2_{02}=V_{0}$ and ${\dot{\phi}}^2_{01}=V_{0}/4$ is solved by changing the variable to $a={v}^{2/3}$. It turns into the equation $4\dot v=\pm\sqrt{3V_0}(1+2v^2)$, whose solution gives the scale factor
\begin{equation}
a=\left[ \frac{1}{\sqrt{2}}\tan{\frac{\sqrt{6V_{0}}\Delta t}{4}}\right]^{2/3},\label{afinetuning}
\end{equation}
where $\Delta t= t-t_{0}$. Besides, the scalar fields $\phi_1$ and $\phi_2$ are obtained after integrating Eq. (\ref{aphipto.1}),
\begin{eqnarray}
\phi_{1}=\frac{2}{\sqrt{3}}\ln\left[\sin{\frac{\sqrt{6V_{0}}\Delta t}{4}}\right],
\\
\phi_{2}=-\frac{2}{\sqrt{3}}\ln\left[\cos{\frac{\sqrt{6V_{0}}\Delta t}{4}}\right]
\end{eqnarray}
The solution (\ref{afinetuning}) represents a universe starting as it were dust dominated by the  scalar field $\phi_1$ with $a\approx\Delta t^{2/3}$. After that it expands and ends in a big rip singularity at finite time. The big rip is reached where $\phi_2$ blows up. 

Finally using the Eq. (\ref{aphipto.1}) we are able to obtain the potential (\ref{pot.gnral}) 
\begin{equation}
V(\phi_1,\phi_2)=\frac{V_{0}}{4}\left[e^{-\sqrt{3}\phi_1}+3e^{\sqrt{3}\phi_2}\right],\label{gral.pot.2}
\end{equation}
as a sum of two separate potentials, each one depending on separate fields $V(\phi_1,\phi_2)$=$V_1(\phi_1)+V_2(\phi_2)$. This is in agreement with the precursor quintom model proposed in the literature \cite{QuintoTransw}. 

Now, we investigate the degenerate case $\beta=-\sigma$ with $\sigma>0$. It gives rise to singular and bouncing solutions with a final de Sitter behavior. The Friedmann equation (\ref{Hdea}) reads 
\begin{equation} 
3H^{2}=V_{0}+ \frac{\sigma+1}{2}\left({\dot{\phi}}^2_{01}-{\dot{\phi}}^2_{02}\right)a^{-\frac{6}{1+\sigma}}.\label{Hdea.ej.2}
\end{equation}

Here there is a degeneration because the contribution of both fields are proportional. When $\dot\phi^2_{01}>\dot{\phi}^2_{02}$, we recover the $\Lambda$CDM cosmological model for $\sigma=1$ and cosmic string for $\sigma=2$. When $\dot\phi^2_{01}<\dot{\phi}^2_{02}$, the solution bounces where the total energy density vanishes. Solving the Friedmann equation (\ref{Hdea.ej.2}), we obtain the solutions 
\begin{eqnarray}
a^+=\left[\sqrt{b}\sinh{\omega \Delta t}\right]^{\frac{\sigma+1}{3}}, \qquad b>0 \label{atsoluc.2},\\
a^-=\left[\sqrt{-b}\cosh{\omega \Delta t}\right]^{\frac{\sigma+1}{3}}, \qquad b<0 \label{atsoluc.2'},
\end{eqnarray}
where $\omega^{2}=3V_{0}/(1+\sigma)^{2}$ and $b=(\sigma+1)(\dot\phi^2_{01} - \dot\phi^2_{02})/2V_{0}$. The scale factor (\ref{atsoluc.2}) evolves like $a\propto t^{(\sigma+1)/3}$ near the singularity, having an inflationary phase for $\sigma>2$ and ending in a de Sitter stage. The bouncing solution (\ref{atsoluc.2'}) begins and ends with de Sitter phases. From Eqs. (\ref{aphipto.1}), (\ref{atsoluc.2})-(\ref{atsoluc.2'}), after integrating, we find the fields in terms of the cosmic time
 {\begin{eqnarray}
\phi_1^+=\dot\phi_{01}\phi^+, \qquad \phi_2=\dot\phi_{02}\phi^+,\\
\phi_1^-=\dot\phi_{01}\phi^-, \qquad \phi_2=\dot\phi_{02}\phi^-,
\end{eqnarray}
with
\be
\phi^+=\frac{1}{\omega\sqrt{b}}\ln{\tanh{\frac{\omega \Delta t}{2}}},
\quad
\phi^-=\frac{1}{\omega\sqrt{-b}}\tan^{-1}{e^{\omega\Delta t}}.
\ee
Hence, coming back to the Eq. (\ref{pot.gnral}) we get the potential corresponding to either case, quintessence or phantom dominated models
\begin{eqnarray}
V=V_{0}\left[ 1 +\frac{\sigma }{1+\sigma}\sinh^{2}{\sqrt{b}\,\omega\phi^+}\right] \label{soluc.pot.2},
\\
V=V_{0}\left[ 1 -\frac{\sigma}{1+\sigma}\sin^{2}{\sqrt{-b}\,\omega\phi^-}\right] \label{soluc.pot.2b}.
\end{eqnarray}
In this models there is no transition from non-phantom to phantom behavior because the 
barotropic index has a definite sign.

\subsection{$3$-dimensional internal space}

Now, we study the case of a multi-scalar field model containing three fields with the following internal metric   
\begin{equation} 
Q_{ik}=
\begin{pmatrix}
1&0&0\\
0&1&0\\
0&0&-1\\
\end{pmatrix}.
\end{equation}
It represents a scalar field configuration where the dimension of the internal space associated with the quintessence sector $N_q$ exceeds the dimension of the respective phantom sector $N_{ph}$, so $N_{q}>N_{ph}$.  The case where the phantom dominates over quintessence components $N_{ph}>N_{q}$ can be obtained from the latter by changing the sign of the metric $Q_{ik}\rightarrow -Q_{ik}$. 

We focus on the case where the multi-scalar field is driven by the potential
\begin{equation}
V(\phi_{1},\phi_{2},\phi_3)=V_0+\frac{1}{2}\dot\phi^2_2-\frac{1}{2}\dot\phi^2_3,\label{3pot}
\end{equation}
so the Friedmann equation reads
\be
\n{300}
3H^2=V_0+\frac{\dot\phi^2_{10}}{a^6}+\frac{\dot\phi^2_{20}}{a^3}-\frac{\dot\phi^2_{30}}{a^3}.
\ee
The four terms of the total energy density can be arranged as a two-fluid mixture, with positive energy densities, that are conserved separately
\ben
\n{3r}
&&\rho_b=\frac{\dot\phi^2_{20}}{a^3},\qquad  \rho_{de}=V_0+\frac{\dot\phi^2_{10}}{a^6}-\frac{\dot\phi^2_{30}}{a^3},\\
&&\dot{\rho}_{b} + 3H{\rho}_{b} =0, \qquad \dot{\rho}_{de} + 3H\gamma_{de}{\rho}_{de}=0,\\
&&\gamma_{de}=\frac{2\dot\phi^2_{10}-\dot\phi^2_{30}a^3}{\dot\phi^2_{10}-\dot\phi^2_{30}a^3+V_0a^6}.
\een
and $\dot\phi^4_{30}\leq4V_{0}\dot\phi^2_{10}$. We have identified $\rho_b$ and $\rho_{de}$  with the baryonic and dark energy densities. The latter component comprises vacuum energy density $V_0$, a stiff fluid term $\rho_s=\dot\phi^2_{10}/ a^{6}$ and a kind of pressurless perfect fluid with negative energy density $\rho_D=-\dot\phi^2_{30}/a^3$. The last term mimics the negative part of the classical Dirac Field \cite{CataChime:2007}. Hence, the  internal space permits to incorporate the negative part of the classical Dirac Field as a source of Einstein equation in a natural way.

The general solution of the Friedmann equation (\ref{300}) takes the following form
\begin{equation}
a^3=\sqrt{\frac{\dot\phi^2_{10}}{V_0}}\sinh{\sqrt{3V_0} t} + \frac{\dot\phi^2_{20}-\dot\phi^2_{30}}{2V_0}\left[\cosh{\sqrt{3V_0} t}-1\right],\label{atsoluc.3}
\end{equation}
where the initial singularity was fixed at $t=0$. The dark energy crosses the phantom divide at $a_{c}=(2\dot\phi^2_{10}/\dot\phi^2_{30})^{1/3}$ where it reaches its minimum value $\rho_{dec}=V_0-\dot\phi^4_{30}/4\dot\phi^2_{10}$.

\begin{figure}[!ht]
\centering
\includegraphics[height=6cm, width=8cm]{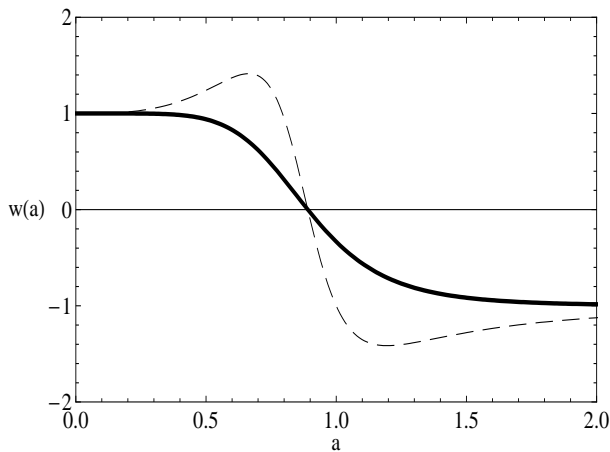}
\caption{The state parameter $w_{de}$ as function of the scale factor. The dashed line includes the Dirac term and the solid line does not include it. }
\end{figure}
\no Basically, the dark energy crosses the  phantom divide due to the term  $\dot\phi^2_{30}$ \cite{CataChime:2007}. 
Concerning the nature of this parameter, note that the dominant energy condition is violated in this cosmological scenario thanks to the term $\dot\phi^2_{30}$. When it is absent the dark energy state parameter $w_{de}$ always remains in the range $[-1,1]$ (see Fig.3), implying  that the dark energy component satisfies the dominant energy condition.

\section{Summary and Conclusion}

We have extended the quintom model by introducing a multi-scalar field configuration along with a $N$-dimensional internal space endowed with an constant internal metric and given the lagrangian formulation of this new model. This formulation enriches the duality between contracting and expanding cosmologies. We have shown that power law solutions are asymptotically stable when the multi-scalar field is driven by the exponential potential with the same internal symmetry than the kinetic energy term, generalizing similar results obtained for the quintessence cosmology. 

We have found the general solution of the nonlinear Einstein-Klein-Gordon equations for the exponential potential $V=V_0e^{-\balpha\cdot \bphi}$, extending previous works with standard scalar fields \cite{Soluc.Exacta} and k-essence \cite{ks}. The richness introduced by the $N$-dimensional internal space and the internal metric $Q_{ik}$ leads, in a natural way, to new singular solutions with a future big rip. For instance, in the explicit solution case, the scale factor blows up at finite time for $\balpha^2<0$, exhibiting a superaccelerated behavior and ending in a final big rip. It has a resemblance with the big rip behavior in the phantom universe reported recently by Wei \cite{Weibigrip:2005} for the potential $V=V_{0}\Phi^{-1}$.

We have shown that the quintom model with a separable potential can be interpreted as a system of coupled fluids. This framework  contains the $\Lambda$CDM model, conduces to bouncing universes and mimics  the cosmic string cosmological model with cosmological constant. 

To gain insight into the internal space structure we have studied a model with a metric $Q_{ik}$, representing a configuration where the dimension of the internal space associated with the quintessence sector $N_q$ exceeds the dimension of the respective phantom sector $N_{ph}$. Choosing $(N_q,N_{ph})=(2,1)$ we have accommodated a 3-scalar field with the following three components, stiff fluid, dust and a dust with negative energy density. Then, the energy density of the dark energy, composed with the latter component, stiff fluid and vacuum energy crosses the phantom divide. Comparing with the Ref.\cite{CataChime:2007} we see that the dust component with negative energy density plays the same role that the negative part of a classical Dirac Field.

\section*{Acknowledgments}

The authors acknowledge the partial support of the University of Buenos Aires under project X044. LPC acknowledges the support of the Consejo Nacional de Investigaciones Cient\'{\i}ficas y T\'ecnicas (CONICET) under project 5169. MGR acknowledges the support of the Consejo Nacional de Investigaciones Cient\'{\i}ficas y T\'ecnicas (CONICET).
R.L. is supported by
the University of the Basque Country through research grant
GIU06/37 and by the former Spanish Ministry of Education and
Culture through research grant FIS2007-61800.


\end{document}